\documentclass[twocolumn,aps]{revtex4}
\usepackage{graphicx}
\usepackage{amsmath}
\usepackage{amssymb}
\usepackage{wasysym}

\begin{document}

\title{Codimension-two points in annular electroconvection as a function of aspect ratio}
\author{V.B. Deyirmenjian$^1$, Zahir A. Daya$^{1,2}$, and Stephen W.
Morris$^1$}
\affiliation{$^1$Department of Physics, University of
Toronto, 60 St. George St., Toronto, Ontario, Canada M5S 1A7\\
$^2$Defence R\&D Canada - Atlantic, 9 Grove Street, PO Box 1012, Dartmouth, Nova Scotia, Canada B2Y 3Z7}
\date{\today}

\begin{abstract}
We rigorously derive from first principles the generic Landau amplitude equation that describes the primary bifurcation in electrically driven convection. Our model accurately represents the experimental system: a weakly conducting, submicron thick liquid crystal film suspended between concentric circular electrodes and driven by an applied voltage between its inner and outer edges. We explicitly calculate the coefficient $g$ of the leading cubic nonlinearity and systematically study its dependence on the system's geometrical and material parameters. The radius ratio $\alpha$ quantifies the film's geometry while a dimensionless group ${\cal P}$, similar to the Prandtl number, fixes the ratio of the fluid's electrical and viscous relaxation  times. Our calculations show that for fixed $\alpha$, $g$ is a decreasing function of ${\cal P}$, as ${\cal P}$ becomes smaller, and is nearly constant for ${\cal P} \gtrsim 1$. As ${\cal P} \rightarrow 0$, $g \rightarrow \infty$. We find that $g$ is a nontrivial and discontinuous function of $\alpha$. We show that the discontinuities occur at codimension-two points that are accessed by varying $\alpha$.

\end{abstract}

\pacs{}

\maketitle

\section{\label{introduce}Introduction}

With the substantial progress realized in the field of pattern formation over the last $15$ years has come an increased need to make stricter comparisons between experiments and their theoretical descriptions \cite{bodenschatz_00}. It has become increasingly important for first principles theories and numerical models to closely mirror experimental systems in order to make unambiguous comparisons. A broad range of phenomena is exhibited by laboratory pattern-forming systems, for example, stationary and traveling patterns, spiral defect chaos, localized structures, and so on.   An experimental system is typically specified by several dimensionless control parameters which span regions where different patterns are observed. These regions are bounded by lines or planes in parameter space.  Codimension-two (CoD2) points occur at special values of the control parameters, and are the nonequilibrium analogs of multicritical points.  Near such points, especially interesting and complex pattern interactions may be expected \cite{cod2}.

One of the most successful approaches used to study patterns is the Landau amplitude equation for pattern amplitudes near a bifurcation. Since the equation can be deduced from symmetry, it has found broad applicability in several experimental systems including Rayleigh-B{\'e}nard convection (RBC), Taylor vortex flow (TVF), and electrohydrodynamic convection in nematic liquid crystals (EHC) \cite{ch93}. In this paper, we apply the Landau amplitude  formalism to thin film electroconvection in an annular geometry, a system that has previously been studied both experimentally\cite{DMjdeb97,annular98,PRE01,PRE02,TDM03} and theoretically at the linear stability level\cite{DDM_pof_99}. This system has a rich bifurcation behavior, with numerous CoD2 points\cite{PRE02,LR98}.

The Landau amplitude equation can be rigorously derived from the complete set of underlying dynamical equations by perturbative expansions about the bifurcation point. Very different physical systems with a common symmetry-breaking bifurcation have, up to parameter-dependent coefficients, identical amplitude equations. Whereas this universal description has been tremendously successful in describing the pattern near the bifurcation, its quantitative verification relies heavily on comparing absolutely the parameter-dependencies of the measured and calculated coefficients. This strategy has seldom been executed and to our knowledge only in RBC, TVF, and EHC has it been generally successful \cite{ch93}. Whereas these systems are three-dimensional (3D), electroconvection in an annular fluid film, the system discussed here, is two-dimensional (2D). RBC, TVF, and to a lesser extent EHC, in large part owe their spatio-temporal richness to the extra spatial dimension. The comparatively simpler spatio-temporal structure of 2D thin film electroconvection is accentuated by its special geometry. The annular film results in a naturally periodic experimental system. This azimuthal periodicity, and thus the absence of lateral boundaries, simplifies the theoretical treatment and invites interesting experimental scenarios such as interposing convection with shear in a closed channel \cite{annular98,DDM_pof_99,PRE02}.

\begin{figure}
\includegraphics[height=4cm]{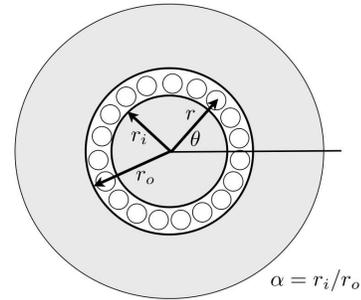}
\caption{\label{geometry}
The coordinate system and film geometry.}
\vspace{0.0cm}
\end{figure}

In this paper, we present a detailed study of the primary bifurcation to electroconvection in an annular film. The film geometry is shown in Fig.~\ref{geometry}. We have previously shown experimentally that the system is adequately modeled by a Landau amplitude equation with a cubic nonlinearity \cite{PRE01}. Our aim here is to rigorously derive the parameter-dependence of the coefficient $g$ of the cubic nonlinearity and compare it with existing measurements from experimental data. In Section~\ref{previous}, we introduce the experimental system, its systematically variable dimensionless parameters, and summarize the protocol by which $g$ was measured. The theoretical model comprising the underlying physics is briefly introduced in Section~\ref{formalize}. We proceed to set up the mathematical formalism to reduce the more complicated basic equations to the Landau amplitude equation truncated at cubic order in the perturbation expansion. We relegate details of the multiple-scales expansion to the Appendix. We study how $g$ varies with the two dimensionless parameters and compare the trends with experiments in Section~\ref{discuss}. We discuss the implications of our work and present a brief conclusion in Section~\ref{conclude}. 

\subsection{\label{previous}Previous experimental results}

Electroconvection in an annular film has been the subject of several  experimental and theoretical studies which have examined convection patterns near onset, interaction with shear flows, linear stability, bifurcations near CoD2 points, and more recently turbulent convection
~\cite{annular98,DDM_pof_99,PRE01,PRE02,TDM03}.

The electroconvection cell is shown schematically in Fig.~\ref{geometry}. It consists of an annulus bordered by two concentric stainless steel electrodes with inner (outer) radii $r_i$ ($r_o$) $\sim 1$~cm. A film of smectic A octylcyanobiphenyl (8CB) liquid crystal doped with tetracyanoquinodimethane (TCNQ) spans the annulus. The resulting film is a weakly conducting 2D annular disk of width $d=r_o-r_i$, radius ratio $\alpha=r_i/r_o$, and thickness $s \sim 0.2~\mu$m. Thickness inhomogeneities relax in the freely suspended film which retains its thickness uniformity even when convecting. The fluid response is Newtonian and the material parameters of the liquid crystal are well characterized by its 2D mass density $\rho$, molecular viscosity $\eta$, and electrical conductivity $\sigma$.  The cell is housed in a vacuum chamber which doubles as a Faraday cage. An experiment consists of drawing a uniform film, placing it under a vacuum, and applying a dc voltage $V$ to the inner electrode while holding the outer electrode at ground potential. The current $I$ through the film is measured. By varying $V$ a current-voltage characteristic is obtained. More details regarding the experimental apparatus and the data acquisition procedure are given in Ref.~\cite{PRE01}. 

\begin{figure}
\includegraphics[height=8cm]{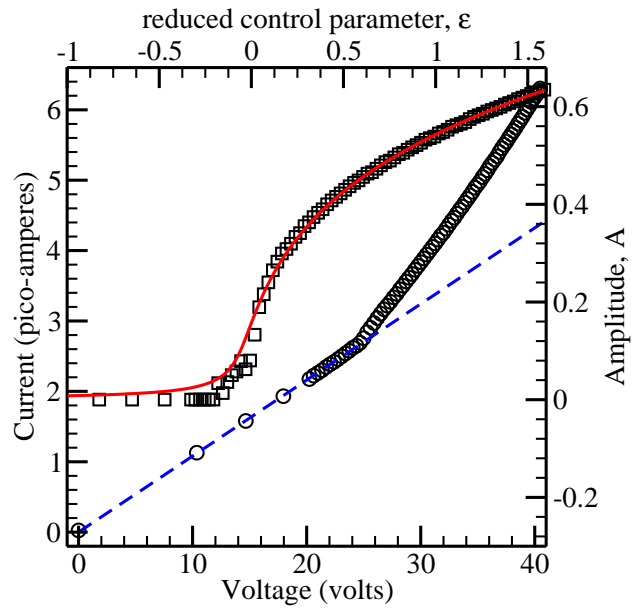}
\caption{\label{iv_ae}
(Color online) A representative current-voltage characteristic $(\circ)$, the corresponding amplitude-$\epsilon$ curve $(\Box)$, and the fit to the Landau amplitude equation (solid line). In this case, we have $\alpha = 0.47$, ${\cal P} =20.4$, and best-fit parameters $g=1.9 \pm 0.1$, $h=5.1 \pm 0.2$, and $f=0.008 \pm 0.002$. The dashed line is a linear fit to the current-voltage data for $V<V_c$, indicating the region of ohmic response.  At the onset of electroconvection, the data depart from the ohmic response.}
\vspace{0.0cm}
\end{figure}

A representative $I-V$ curve is shown in Fig.~\ref{iv_ae}. A convection threshold at a critical voltage $V=V_c$ is clearly observed. The current is transported by ohmic conduction for $V<V_c$ while convection contributes for $V>V_c$. An experimental realization is categorized by the dimensionless parameters $\alpha$, already introduced, and ${\cal P} = \epsilon_0 \eta/\rho \sigma d$, where $\epsilon_0$ is the dielectric permitivity of free space. The parameter ${\cal P}$ is the ratio of time scales of electrical and viscous processes of the film and is analogous to the familiar Prandtl number of RBC. The experiments explored various $\alpha$ by using different combinations of inner and outer electrodes. At each $\alpha$, several films of different $s$ and consequently different ${\cal P}$ were investigated. 

The $I-V$ data can be expressed in terms of the reduced amplitude $A$ and the forcing parameter  $\epsilon$,
\begin{equation}
\label{ivtoae}
 A^2 =  \frac{I}{I_{cond}}-1 {\rm ~~and~~} \epsilon = \biggl(\frac{V}{V_c}\biggr)^2-1\,,
\end{equation}
where $I_{cond}$ is the contribution to the total current $I$ that is due to ohmic conduction. Below the onset of convection, $I=I_{cond}$ and $A=0$. When $V>V_c$, then $I>I_{cond}$. This results in $\epsilon >0 $ and $A>0$. The amplitude $A$ and control parameter $\epsilon$ are then effectively an order parameter and a reduced temperature for electroconvection. Having transformed the $I-V$ data using Eq.~\ref{ivtoae}, the $A-\epsilon$ data is then modeled with the phenomenological steady state Landau amplitude equation
\begin{equation}
\label{landau_data}
\epsilon A - gA^3 -hA^5+f=0\,.
\end{equation}
In the modeling, $g$, $h$, and $f$ are fit parameters with $f$ restricted to be positive. Since $g$ can be negative, the quintic term in the amplitude equation is necessary. Details regarding the data modeling have been reported in Ref.~\cite{PRE01}.

Our interest in this paper is to investigate the variation of $g$ with $\alpha$ and ${\cal P}$. Previous experiments~\cite{PRE01} have explored the regimes $0.33 \leq \alpha \leq 0.80$ and $1 < {\cal P} < 150$. Measurements of $g$ were found to be roughly independent of ${\cal P}$ except at $\alpha=0.33$ and $2 < {\cal P}< 8$, where $g$ was seen to increase with increasing ${\cal P}$. As a function of $\alpha$, $g$ was found to be generally increasing with increasing $\alpha$ in overall trend. In the next section we recount features of the physical model and set up the mathematical formalism for perturbatively solving the equations.

\section{\label{formalize}Mathematical Formalism}

A linear instability mechanism gives rise to electroconvection in a freely suspended fluid film \cite{DMjdeb97,DDM_pof_99}. The film, an annular disk on the $z=0$ plane, is subjected to electric potential boundary conditions of $V$ volts at its inner edge ($r=r_i$) and zero volts at its outer edge ($r=r_o$). The inner electrode $r<r_i$ is at $V$ volts and the outer electrode $r>r_o$ is at zero volts. The potential is zero at infinity. The electrostatic boundary value problem prescribed by these conditions, with the film being a conducting liquid, implies that a surface charge density develops on the film's upper and lower free surfaces. Positive charge preferentially accumulates close to the inner positive electrode and negative charge at the grounded outer electrode. Consequently, an electric force is exerted on the fluid by the action of the radially outward component of the electric field on the surface charge density. Note the striking analogy with RBC where thermal boundary conditions on a thermally conducting liquid lead to hotter, less dense fluid near the hot boundary and colder, more dense fluid near the cold boundary. The action of the gravitational field on the mass density exerts a force on the fluid. In both cases when the forcing overcomes dissipation, the fluid becomes linearly unstable to perturbations, resulting in convection. 

The relationship between the electric potential and the surface charge density at any spatial position on the film is nonlocal in electroconvection, unlike thermal convection where the temperature and mass density are locally related. The surface charge density is directly related to the discontinuity in the component of the electric field perpendicular to the film at $z=0$. As a result, electric fields in the full 3D space determine the charge density of the film at $z=0$. 

We use the cylindrical coordinate system $(r,\theta,z)$. The film is in the $z=0$ plane with radial coordinates between $r_i \leq r \leq r_o$ and has the areal material parameters described in Sec.~\ref{previous}. It is assumed that the fluid is uniform in temperature and there is no significant ohmic heating. Since the velocity field is 2D as the film flows in the $z=0$ plane, we choose to describe it implicitly with the streamfunction $\psi(r,\theta)$. The surface charge density is $q(r,\theta)$ and the electric potential is denoted by $\Psi_3(r,\theta,z)$. The electric potential in the film is for convenience denoted $\Psi(r,\theta) = \Psi_3(r,\theta,z=0)$. The fluid is described by the incompressible Navier-Stokes equation with an electrical body force. The surface charge is advected by the flow and transported by ohmic conduction. Laplace's equation is obeyed by the electric potential for $z\neq 0$. Nondimensionalizing lengths by the film width $d=r_o-r_i$, electric potential by $V$, and time by $\epsilon_0 d/\sigma$ in the momentum, charge conservation, and Laplace equations we get the set of governing equations

\begin{eqnarray}
\nonumber \Biggl[ \nabla^2 &-& \frac{1}{{\cal P}}\frac{\partial}{\partial t}\Biggr]
(\nabla \times \nabla \times {\vec \phi}) + {\cal R}
\biggl(\nabla\Psi \times \nabla q \biggr)\\ 
&=&\frac{1}{{\cal P}} \biggl( (\nabla \times {\vec \phi}) \cdot
 \nabla \biggr) ( \nabla \times \nabla \times {\vec \phi}) \,,
\label{NDNS}
\end{eqnarray}
\begin{equation}
\frac{\partial q}{\partial t} + ( \nabla \times {\vec \phi} )
\cdot \nabla q - \nabla^2 \Psi = 0 \,, \label{NDCC}
\end{equation}
\begin{eqnarray}
\nabla_3^2 \Psi_3 &=& 0 \,, \label{NDLAP} \\
q &=& -2 \frac{\partial \Psi_3}{\partial z}{\Biggl|}_{z=0^+} \,,
\label{NDqdef}
\end{eqnarray}
where the dimensionless parameters are
\begin{equation}
{\cal R} \equiv \frac{\epsilon_0^2 V^2}{\sigma \eta} \hspace{5mm} {\rm and}
\hspace{5mm}
{\cal P} \equiv
\frac{\epsilon_0 \eta}{\rho \sigma d} \,. \label{Rayleigh}
\end{equation}
Details regarding the assumptions and the derivation can be found in Refs.~\cite{DMjdeb97,DDM_pof_99}. 

In the standard manner~\cite{ch93}, we write the streamfunction, charge density, and electric potential as the sum of the base state solution and a perturbation. See Ref.~\cite{dey_g_paper} for details. The resulting equations can be succinctly expressed as 

\begin{equation}
{\cal L}{\cal C} = {\cal B} \;. \label{diropeqn}
\end{equation}
Here we have the operator
\begin{equation}
{\cal L} = \left(
\begin{array}{cccc}
{\nabla}^4 &
-\frac{{\cal R}}{r}{\partial_r}\Psi^{(0)}{\partial_\theta} &
\frac{{\cal R}}{r}{\partial_r}q^{(0)}{\partial_\theta} & 0 \\
-\frac{1}{r}{\partial_r}q^{(0)}{\partial_\theta} &
0 & {\nabla}^2 & 0 \\
 0 & 1 & 0 & 2\partial_z(.)|_{z=0^+} \\
 0   & 0 & 0 & \nabla_3^2 \\
\end{array}\label{directop}
\right) \;.
\end{equation}
We have denoted functions of the base state by superscript $(0)$. Note that ${\nabla^2} = {\partial_{rr}}+ \frac{1}{r}{\partial_r} +\frac{1}{r^2}{\partial_{\theta\theta}} $ while $\nabla_3^2= {\nabla^2}+{\partial_{zz}}$. The functions ${\cal C}$ and ${\cal B}$ are

\begin{equation}
{\cal C} = \left(
\begin{array}{c}
\phi(r,\theta) \\
q(r,\theta) \\
\Psi(r,\theta) \\
\Psi_3(r,\theta,z) \\
\end{array}
\right) \;  \label{directC}
\end{equation}
and 
\begin{equation}
{\cal B} =
 \left(
\begin{array}{c}
[\frac{1}{{\cal P}}\nabla^2\partial_t\phi
-\frac{1}{r{\cal P}}[({\partial_r}\phi)({\partial_\theta}\nabla^2\phi)
        -({\partial_\theta}\phi)({\partial_r}\nabla^2\phi)]\\
        ~~~~~~~~~~~~~+\frac{{\cal R}}{r}[({\partial_\theta}q)({\partial_r}\Psi) - ({\partial_r}q)({\partial_\theta}\Psi)]] \\
        \\
\partial_t q + \frac{1}{r}[({\partial_r}q)({\partial_\theta}\phi) - ({\partial_\theta}q)({\partial_r}\phi)] \\
\\
0 \\
\\
0 \\
\end{array}
\right) \;. \label{B}
\end{equation}

The multiple-scales perturbation theory employed in our treatment is the 
same as that given in Ref.~\cite{dey_g_paper} for electroconvection in a
rectangular geometry. We expand Eq.~\ref{directop} using the slow time scale $T=\epsilon t$, where $\epsilon = {\cal R}/{\cal R}_c-1$ is the reduced control parameter defined earlier in Eq.~\ref{ivtoae}. Collecting terms of the same expansion order in $\epsilon$ we write 

\begin{eqnarray}
\nonumber {\cal L}_0 {\cal C}_0&=&{\cal B}_0 \,,\\
\nonumber{\cal L}_0 {\cal C}_1+{\cal L}_1 {\cal C}_0&=&{\cal B}_1 \,,\\
{\cal L}_0 {\cal C}_2+{\cal L}_1 {\cal C}_1+{\cal L}_2 {\cal C}_0&=&{\cal B}_2 \,,
\end{eqnarray}
for orders $\epsilon^{1/2}$, $\epsilon$, and $\epsilon^{3/2}$. A systematic sequential solution of the above equations results in the necessary condition
\begin{equation}
F_1 \partial_T A + F_2 A + F_4 A|A|^2=0\,.
\end{equation}   
The amplitude equation in the fast variables is of the Landau form
\begin{eqnarray}
\tau \partial_t A & = & \epsilon A -gA|A|^2 \;, \label{ampleq}
\end{eqnarray}
where 
\begin{equation}
\tau = -\frac{F_1}{F_2} {\rm ~~and~~} g= -\frac{F_4}{F_2} \,.
\end{equation}
The functions $F_1$, $F_2$, and $F_4$ are further discussed in the Appendix. Other details regarding the numerical evaluation of $g$ and $\tau$ are also given in the Appendix.

\section{\label{discuss}Discussion}

\begin{figure}
\includegraphics[height=7cm]{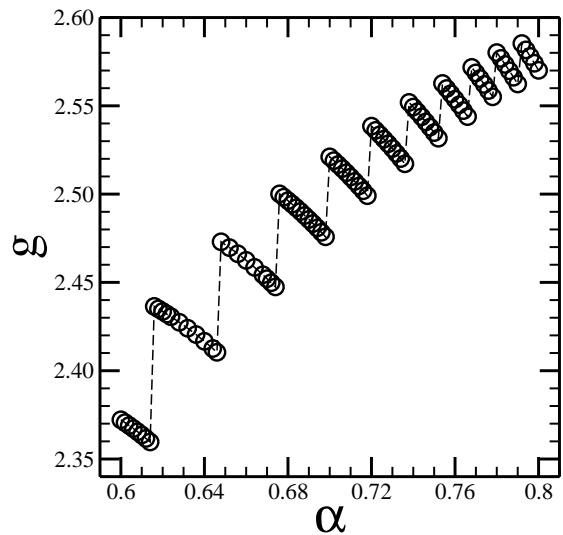}
\caption{\label{g_vs_rr}
$g$ versus $\alpha$ for ${\cal P}=123$.
}
%\vspace{0.0cm}
\end{figure}

In this section we present the results of our calculations of the functional dependencies of $g = g(\alpha,{\cal P})$. Since the experiments on electroconvection are, due to the values of the physical and dimensional parameters, constrained to large ${\cal P} > 1$, we start by discussing $g = g(\alpha)$ for large and essentially infinite ${\cal P}$. In figure~\ref{g_vs_rr} is plotted the calculated values of $g$ for $0.6 \leq \alpha \leq 0.8$. In overall trend, the Landau cubic coefficient $g$ increases with $\alpha$ and approaches a limiting value as $\alpha \rightarrow 1$. At 
$\alpha =0.8$, $g=2.570$ is within $10\%$ of the limiting value $g=2.842$ calculated earlier \cite{dey_g_paper} in a Cartesian or rectangular geometry. The coefficient $g(\alpha)$ decreases with $\alpha$ for intervals over which the critical mode $m=m_c^0$ is constant. This trend is punctuated by discontinuities in $g$ at values of $\alpha$ where both $m$ and $m+1$ are equally unstable; these are the  codimension-two (CoD2) points. At these jumps, the value of $g$ increases and more than compensates for the region of decrease at each $m$. The result is an overall increasing trend.

Table~\ref{gvsrrtable} compares experimental measurements of $g$ from Ref.~\cite{PRE01} at six radius ratios with the results of the present calculations. First note that in spite of the large scatter in the experimental measurements, they still show the overall increasing trend of $g$ with $\alpha$. Further, the experiments have widely separated values of $\alpha$ and so are not able to resolve the discontinuities in $g$ for which further experimental work in a restricted but densely sampled range of $\alpha$ would be needed. And finally, in comparing measurements to the calculations, there is a significant disparity on the order of $10-30\%$. Interestingly, the experimental measurements show that as the radius ratio is decreased, the Landau cubic coefficient becomes negative. This implies that there is at least one tricritical point ($g=0$) that demarcates the super- and subcritical branches. The current calculations are cumbersome to extend to smaller $\alpha$ because this requires additional orders in the expansion of the stream function\cite{dey_unpublished}. 

The CoD2 discontinuities in $g$ are larger at smaller $\alpha$ as is evident from Fig.~\ref{g_vs_rr}.  To further quantify this observation, we have graphed the fractional discontinuity in $g$, $2(g_{m+1}-g_m)/(g_{m+1}+g_m)$, versus the fractional change in the mode number, $2(m+1-m)/(m+1+m)$, at several CoD2 points, as shown in Fig.~\ref{jumps}. In the above, $g_m$ is the value of $g$ for a particular critical mode $m=m_c^0$. For the CoD2 point at $m=8,9$, the jump in $g$ is about $3\%$ suggesting that experiments looking at the behavior near CoD2 points will have typically to measure $g$ with $1\%$ resolution. The current resolution of experimental measurements is about $10\%$ and will have to be significantly improved if any meaningful study is to be made. On the other hand, since the fractional discontinuity in $g$ increases strongly with the fractional discontinuity in mode number, it is conceivable that at small mode CoD2 points, say $m=3,4$, the discontinuity in $g$ will be very large and measurements with $5-10\%$ accuracy may suffice. This will involve working at small $\alpha$ which is a challenge experimentally since broad films will have to be drawn.

\begin{table}[t]
\center
\begin{tabular}{|c|c|c|c|}\hline
 $\alpha$&$g$ (experiment)& ${\cal P}$ &$g$ (theory) \\ \hline
0.33&-0.74 $\pm$ 0.23&$2.1<{\cal P}<4.4$& \\
0.47&~1.64 $\pm$ 0.06&$13.5<{\cal P}<20.7$& \\
0.56&~0.73 $\pm$ 0.15&$59.4<{\cal P}<100.8$& \\
0.60&~2.72 $\pm$ 0.34&$31.3<{\cal P}<38.9$& 2.372\\
0.64&~1.87 $\pm$ 0.10&$25.2<{\cal P}<63.0$& 2.417\\
0.80&~2.21 $\pm$ 0.29&$15.3<{\cal P}<142.8$& 2.570 \\  \hline
1.00~(`plate')&~&${\cal P} = \infty$&2.842 \\  \hline
\end{tabular}
\vskip 0.15in
\caption{Experimental measurements of the coefficient of the cubic nonlinearity, $g$. The theoretical $g$ at $\alpha=0.60$, $0.64$, and $0.80$ are for ${\cal P}=123$.}
\label{gvsrrtable}
\vspace{2mm}
\end{table}

\begin{figure}
\includegraphics[height=7cm]{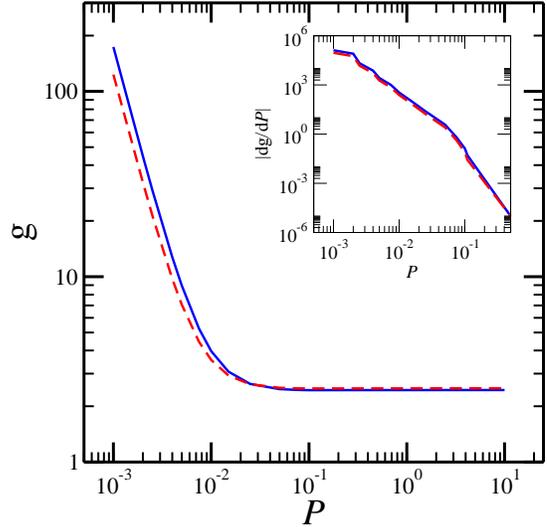}
\caption{\label{g_vs_prandtl}
(Color online) The coefficient $g$ versus ${\cal P}$ at $\alpha=0.674$ (solid line) and $\alpha=0.676$~(dashed line). 
The inset shows $|dg/d{\cal P}|$ versus ${\cal P}$ for the same $\alpha$.
}
\vspace{0.0cm}
\end{figure}

We now discuss the ${\cal P}$-dependence of $g$ with constant $\alpha$. The calculated values of $g$ for $0.001 \leq {\cal P} \leq 10$ for $\alpha = 0.674$ and $0.676$ are plotted in Fig.~\ref{g_vs_prandtl}. These $\alpha$ 
straddle the CoD2 point for $m=10,11$. The cubic Landau coefficient is practically constant for ${\cal P} > 0.1$. However, for small ${\cal P}$, $g$ diverges as ${\cal P} \rightarrow 0$. The inset in Fig.~\ref{g_vs_prandtl} 
shows the absolute value of $dg/d{\cal P}$ as a function of ${\cal P}$. As ${\cal P}$ decreases, the discontinuity in $g$ at the CoD2 point $m=10,11$ increases as is seen by the two diverging curves in Fig.~\ref{g_vs_prandtl}. 
For ${\cal P} > 1$, the fractional discontinuity is about $2\%$ growing to $10\%$ for ${\cal P} = 0.01$ and $30\%$ for ${\cal P} = 0.001$. This suggests that experimental measurements near CoD2 points are better performed at small 
${\cal P}$.  The combination of small $\alpha$ and ${\cal P}$ are required for large changes in $g$ through CoD2 points. Unfortunately, experiments can seldom access this regime of parameter space. Most experiments have been performed for $\alpha > 0.30$ and ${\cal P} > 1$. The latter restriction is primarily because the material parameters are only slightly adjustable.  

\begin{figure}
\includegraphics[height=7cm]{DDM04_FIG4.eps}
\caption{\label{jumps}
The fractional discontinuity in $g$, $2\Delta g/(g_m+g_{m+1})$, versus the fractional change in the critical mode number $m=m_c^0$, $2\Delta m/(2m+1)$, at CoD2 points with ${\cal P}=123$. }
\vspace{0.0cm}
\end{figure}

Langford and Rusu have considered several qualitative bifurcation scenarios 
for this system \cite{LR98}. They focused on the possible transitions at
CoD2 points. Our work indicates that the bifurcations at CoD2 points are
supercritical for most radius ratios. Subcritical transitions may be possible at small $\alpha$ \cite{dey_unpublished}.

\section{\label{conclude}Conclusion}

In summary, we have examined the variation of $g$, the coefficient of the nonlinear term in the amplitude equation Eq.~\ref{ampleq}, with radius ratio $\alpha$ and Prandtl number ${\cal P}$ for electroconvection in a two-dimensional annular fluid. We have observed that the steady-state amplitude of the bifurcation from a conducting to a convecting fluid can change discontinuously as $\alpha$ passes through CoD2 points. The dependence of $g$ on ${\cal P}$ is not strong, except for ${\cal P}<0.1$. 

%We currently are analyzing the nonlinear dynamics at codimension-two points with smaller $\alpha$ \cite{dey_unpublished}. 

In the experiments considered here, the variation of $g$ is a consequence of changing the geometry of the system via the radius ratio $\alpha$. It is also possible to pass through CoD2 points by applying a Couette shear to the fluid at fixed $\alpha$ \cite{PRE02}. Experiments revealed that $g$ was found generally to decrease with increasing shear and discontinuities in $g$ were observed at CoD2 points \cite{PRE02}. Applying the methods employed in this paper to the case of nonzero shear, it should be possible to establish whether codimension-three (CoD3) points exist at special values of the parameters  \cite{dey_unpublished}.

\appendix*
\section{Multiple-Scales Expansion}
We used a multiple-scales formalism analogous to that developed for RBC in Ref.~\cite{ch93}, in which a compatability condition is derived for the amplitude of the slowly varying envelope of the convection rolls. We generally adhere to the notational conventions of Ref.~\cite{ch93} and our treatment is similar to that of Ref.~\cite{dey_g_paper}, adapted to the annular geometry and including the effects of finite $\cal P$.  The multiple-scales expansion of Eq.~\ref{diropeqn} gives
Eq.~\ref{ampleq} where $\epsilon=({\cal R}-{\cal R}_c)/{\cal R}_c$,
$\tau=-F_1/F_2$, $g=-F_4/F_2$,
\begin{eqnarray}
F_1 & = & \langle \bar q_{b0}^* \bar q_0 + 
  {\cal P}^{-1} \bar\phi_{b0}^*\nabla^2 \bar\phi_0  \rangle\;,\label{F1} \\
F_2 & = & im_c^0 {\cal R}_c^0 \langle
  \frac{\bar\phi_{b0}^*}{r} [{\partial_r}\Psi^{(0)}\bar q_0-{\partial_r}q^{(0)}\bar\Psi_0]
  \rangle \;,\label{F2} \\
F_4 & = & im_c^0 \langle \frac{ \bar q_{b0}^* {\cal R }_c^0}{r}[
    -\bar q_0^*{\partial_r}\Psi_1^{\epsilon} +\bar q_0{\partial_r}\Psi_2^{\epsilon} -2{\partial_r}\bar q_0^*\Psi_1^{\epsilon} 
     \nonumber \\
    & & +2q_1^{\epsilon}{\partial_r}\bar\Psi_0^* +{\partial_r}q_1^{\epsilon}\bar\Psi_0^*-{\partial_r}q_2^{\epsilon}\Psi_0 ]
    -\frac{ \bar\phi_{b0}^* }{r{\cal P}}[  2{\partial_r}\bar\phi_0^*\nabla^2\phi_1^{\epsilon} \nonumber \\
    & &  +\bar\phi_0^*{\partial_r}\nabla^2\phi_1^{\epsilon} -{\partial_r}\phi_1^{\epsilon}\nabla^2\bar\phi_0^* -2\phi_1^{\epsilon}{\partial_r}\nabla^2\bar\phi_0^* ] +\frac{ \bar q_{b0}^* }{r} [ \nonumber \\
    & &  2{\partial_r}\bar q_0^*\phi_1^{\epsilon}
    +\bar q_0^*{\partial_r}\phi_1^{\epsilon} -{\partial_r}q_1^{\epsilon}\bar\phi_0^* +{\partial_r}q_2^{\epsilon}\bar\phi_0 \nonumber \\
& & -2q_2^{\epsilon}{\partial_r}\bar\phi_0^* ] \rangle \;, \label{F4} \\
\langle \ldots \rangle & = & \int_{r_i}^{r_o} r dr \; (\ldots) \;.
\end{eqnarray}
Complex conjugation is represented by a superscript $^*$.
The functions that appear in Eqs.~\ref{F1}-\ref{F4} are as follows.
The solutions of the linear stability problem at $m=m_c^0$ are
\begin{equation}
\begin{array}{lll}
\phi_0 = \bar\phi_0 e^{im_c^0\theta} \; &
q_0 = \bar q_0 e^{im_c^0\theta} \; &
\Psi_0 = \bar\Psi_0 e^{im_c^0\theta} \\
\bar\phi_0=\sum_p \bar A_p\bar\phi_{0p} \; &
\bar q_0=\sum_p \bar A_p\bar q_{0p} \; &
\bar\Psi_0=\sum_p \bar A_p\bar\Psi_{0p}
\end{array} \label{dirsol}
\end{equation}
\begin{eqnarray}
\bar\phi_{0p} & = & C_{m_c^0;p} \;, \label{phibarp} \\
\bar q_{0p} & = & i\sum_l v_{0pl} q_{m_c^0;l} \;, \label{qbarp} \\
\bar\Psi_{0p} & = & i\sum_l v_{0pl} \psi_{m_c^0;l} \;. \label{Psibarp}
\end{eqnarray}
The expansion functions $C_{m;p}(r)$ and $\psi_{m;l}(r,z)$ satisfy the
boundary conditions in polar coordinates\cite{DDM_pof_99} and 
$q_{m;l}(r)=-2[\partial_z\psi_{m;l}(r,z)]|_{z=0^+}$.
The functions
\begin{equation}
\begin{array}{lll}
\phi_1^{\epsilon}=i\sum_p E_p C_{2m_c^0;p} \; & q_1^{\epsilon}=\sum_l a_l q_{2m_c^0;l} \; &
\Psi_1^{\epsilon}=\sum_l a_l \psi_{2m_c^0;l} \\
\phi_2^{\epsilon}=0 \; & q_2^{\epsilon}=\sum_l b_l q_{0;l} \; & \Psi_2^{\epsilon}=\sum_l b_l \psi_{0;l}
\end{array} \label{order1sol}
\end{equation}
\begin{eqnarray}
a_l & = & [\sum_p E_p s_{2m_c^0pl} - \langle N_2 \psi_{2m_c^0;l} 
  \rangle ]/\chi_{2m_c^0l}^2 \;, \\
s_{mpl} & = & \langle \frac{m}{r} {\partial_r}q^{(0)}C_{m;p} \psi_{m;l} \rangle \;, \\
b_l & = & -\langle N_3 \psi_{0;l} \rangle/\chi_{0l}^2 \;, \\
N_1 & = & \frac{m_c^0{\cal R}_c^0}{r}(\bar q_0 {\partial_r}\bar\Psi_0 -
   {\partial_r}\bar q_0 \bar\Psi_0) \nonumber \\
    & & -\frac{m_c^0}{r{\cal P}}({\partial_r}\bar\phi_0\nabla^2\bar\phi_0
          -\bar\phi_0{\partial_r}\nabla^2\bar\phi_0) \;, \\
N_2 & = & \frac{im_c^0}{r}({\partial_r}\bar q_0 \bar\phi_0 -\bar q_0 {\partial_r}\bar\phi_0) \;, \\
N_3 & = & \frac{im_c^0}{r}[
           ({\partial_r}\bar q_0^* \bar\phi_0 - {\partial_r}\bar q_0 \bar\phi_0^*) \nonumber \\
    & & -(\bar q_0 {\partial_r}\bar\phi_0^* - \bar q_0^* {\partial_r}\bar\phi_0) ] \;,
\end{eqnarray}
satisfy the order $\epsilon$ multiple-scales equations. The coefficients
$E_p$ are specified by
\begin{eqnarray}
\sum_p T_{kp} E_p & = & \langle N_1 C_{2m_c^0;k} \rangle  \nonumber \\
  & & - 2m_c^0{\cal R}_c^0 \sum_l \langle N_2 \psi_{2m_c^0;l} \rangle 
   Z_{2m_c^0kl}/\chi_{2m_c^0l}^2 \;, \nonumber \\
  & & \\
T_{kp} & = & \beta_{2m_c^0p}^4\delta_{kp} \nonumber \\
  & &-2m_c^0{\cal R}_c^0\sum_l s_{2m_c^0pl}Z_{2m_c^0kl}/\chi_{2m_c^0l}^2 \;,\\
Z_{mkl} & = & \langle \frac{C_{m;k}}{r}({\partial_r}\Psi^{(0)}q_{m;l}
   -{\partial_r}q^{(0)}\psi_{m;l}) \rangle \;.
\end{eqnarray}
The solutions of the adjoint equations evaluated at $m=m_c^0$ are
\begin{equation}
\begin{array}{lll}
\phi_{b0} = \bar\phi_{b0} e^{im_c^0\theta} \; &
q_{b0} = \bar q_{b0} e^{im_c^0\theta} \; &
\Psi_{b0} = \bar\Psi_{b0} e^{im_c^0\theta} \\
\bar\phi_{b0}=\sum_p B_p\bar\phi_{b0p} \; &
\bar q_{b0} = \sum_p B_p\bar q_{b0p} \; &
\bar\Psi_{b0} = \sum_p B_p\bar\Psi_{b0p}
\end{array} \label{adjsol}
\end{equation}
\begin{eqnarray}
\bar q_{b0p}  & = & \psi_{m_c^0;p} \;, \label{qbarbp} \\
\bar\phi_{b0p} & = & i \sum_l v_{b0pl} C_{m_c^0;l} \;, \label{phibarbp} \\
v_{b0pl} & = & -s_{m_c^0lp}/\beta_{m_c^0l}^4 \;, \\
\bar\Psi_{b0p} & = & -\frac{i}{r}m_c^0{\cal R}_c^0{\partial_r}\Psi^{(0)}
   \bar\phi_{b0p} \label{Psibarbp} \;.
\end{eqnarray}
The coefficients $B_p$ are derived from
\begin{equation}
\sum_p B_p [-\frac{i}{r}m_c^0{\cal R}_c^0{\partial_r}q^{(0)}\bar\phi_{b0p}
  + \nabla^2\bar q_{b0p} +2(\partial_z\bar\Psi_{b0p})|_{z=0^+}] =0 \;.
\end{equation}

The amplitude $A$ is normalized by setting
\begin{equation}
{\rm Nu} -1 = \frac{ \langle\langle q u_r \rangle\rangle }
   { \langle\langle \sigma E_r^{(0)} \rangle\rangle}  \;,
\end{equation}
where 
\begin{equation}
\langle\langle \ldots \rangle\rangle = \frac{1}{2\pi}
 \int_0^{2\pi} d\theta \int_0^{\infty} r dr (\ldots) \;,
\end{equation}
$u_r=\frac{1}{r}\partial_{\theta}\phi$ is the radial component of the velocity field, $\sigma$ is the conductivity, and $E_r^{(0)}$ is the radial component of the base state electric field. The Nusselt number ${\rm Nu}$ is the ratio of the total current density to the conducted current density, spatially averaged \cite{dey_g_paper}.

To find the coefficients of the normalized amplitude equation, Eq.~\ref{ampleq}, we evaluate Eqs.~\ref{F1}-\ref{F4} for a given $\alpha$ and ${\cal P}$. The charge density expansion functions $q_{m;l}$ are computed via an approximation of the electrostatic equations in which $\Psi$ and $q$ are linearly related \cite{DDM_pof_99}. The series solutions of the linear (Eq.~\ref{dirsol}) and adjoint (Eq.~\ref{adjsol}) problems are terminated at $p=1$. The number of expansion functions employed in the linear (Eqs.~\ref{phibarp}-\ref{Psibarp}),
first-order (Eq.~\ref{order1sol}), and adjoint (Eqs.~\ref{qbarbp}-\ref{Psibarbp}) solutions is twenty.

\end{document}